%% file: iclr2025_conference.tex
\documentclass{article} 
\usepackage{iclr2025_conference,times}
\iclrfinalcopy
\input{math_commands.tex}

\usepackage{fontawesome}
\usepackage{hyperref}
\usepackage{url}
\usepackage{graphicx}
\usepackage{booktabs}
\usepackage{multirow}
\usepackage{geometry}
  \title{AudioGen-Omni\faCameraRetro: A Unified Multimodal Diffusion Transformer for
Video-Synchronized Audio, Speech, and Song Generation}

\author{
Le Wang\textsuperscript{1},    
Jun Wang\textsuperscript{2},
Chunyu Qiang\textsuperscript{2},
Feng Deng\textsuperscript{2},
Chen Zhang\textsuperscript{2}, Di Zhang\textsuperscript{2}, Kun Gai\textsuperscript{2}
 \\
\textsuperscript{1} China University of Mining and Technology, 
\textsuperscript{2} Kuaishou Technology \\
\texttt{\{TS23170132P31\}@cumt.edu.cn},
\texttt{\{wangjun06, qiangchunyu, dengfeng\}@kuaishou.com} \\
}

\begin{document}

\maketitle
\begin{abstract}
We present AudioGen-Omni — a unified approach based on multimodal diffusion transformers (MMDit), capable of generating high-fidelity audio, speech, and song coherently synchronized with the input video. AudioGen-Omni  introduces a novel joint training paradigm that seamlessly integrates large-scale video-text-audio corpora, enabling a model capable of generating semantically rich, acoustically diverse audio conditioned on multimodal inputs and adaptable to a wide range of audio generation tasks. AudioGen-Omni employs a unified lyrics-transcription encoder that encodes graphemes and phonemes from both song and spoken inputs into dense frame-level representations. Dense frame-level representations are fused using an AdaLN-based joint attention mechanism enhanced with phase-aligned anisotropic positional infusion (PAAPI), wherein RoPE is selectively applied to temporally structured modalities to ensure precise and robust cross-modal alignment. By unfreezing all modalities and masking missing inputs, AudioGen-Omni mitigates the semantic constraints of text-frozen paradigms, enabling effective cross-modal conditioning. This joint training approach enhances audio quality, semantic alignment, and lip-sync accuracy, while also achieving state-of-the-art results on Text-to-Audio/Speech/Song tasks. With an inference time of 1.91 seconds for 8 seconds of audio, it offers substantial improvements in both efficiency and generality. 

\textbf{Demo: }\url{https://ciyou2.github.io/AudioGen-Omni/}
\end{abstract}

\section{Introduction}
Audio plays a critical role in video content, complementing visual information while reinforcing narrative structure and emotional engagement. In domains such as film production, game design, and social media, components like ambient sound, background music, and voiceover are essential for creating immersive user experiences. Recently, video-to-audio generation tasks—including audio \cite{copet2023simple, kreuk2022audiogen, liu2023audioldm, majumder2024tango, tian2025vidmuse} and speech synthesis \cite{lei2024uni, choi2023diffv2s, ephrat2017vid2speech, le2017generating, le2015reconstructing}—have gained increasing attention \cite{cheng2025mmaudio, tian2025audiox, kim2025faces}, becoming integral to multimedia content creation and demonstrating strong potential in enhancing user experience.

Recent works on video-to-audio and video-to-speech generation have made important strides in modeling cross-modal relationships. For example, Tian et al. \cite{tian2025audiox} propose a dedicated architecture for any-to-audio synthesis that supports a broad range of input modalities—including text, video, image, music, and audio. However, the model does not support speech or singing voice synthesis tasks and exhibits suboptimal alignment between audio effects and rhythmic timing. MMAudio \cite{cheng2025mmaudio} generates synchronized audio from video and/or text via multimodal joint training and a synchronization module for temporal alignment. However, it cannot synthesize speech or singing voice, limiting its scope. DualDub \cite{tian2025dualdub} proposes a unified framework that jointly generates background audio and speech using a multimodal encoder and cross-modal aligner for improved synchronization. However, it lacks explicit lip-sync alignment between speech and video and does not support singing voice generation. Face2Voice \cite{kim2025faces} achieves superior speech synthesis by bridging the modality gap via hierarchical representation learning but lacks support for sound effects or song, and its language naturalness requires improvement. VidMuse \cite{tian2025vidmuse} advances video-to-music generation with a large dataset and effective alignment techniques but does not support speech or singing voice synthesis.

Such approaches typically employ task-specific designs, leading to suboptimal alignment across modalities and reduced generation quality. Integration of multimodal cues—such as lip movements and facial expressions in silent video, prosodic and phonetic characteristics in speech and singing, and ambiguous textual semantics—remains a significant challenge. Moreover, current models generally lack flexible conditioning mechanisms capable of accommodating diverse input combinations, and fail to support the unified generation of various audio types. Consequently, the absence of a comprehensive, general-purpose framework that can synthesize audio, music, and speech within a unified model continues to impede progress in audio-video fusion and multimodal generation research.

To address these challenges, we propose AudioGen-Omni, a unified Multimodal Diffusion Transformer (MMDiT) framework that integrates video, audio, and text modalities within a shared semantic space to enable high-fidelity generation of diverse audio types, including general audio, speech, and song. AudioGen-Omni supports flexible multimodal conditioning and accommodates various generation tasks within a single architecture. We introduce a lightweight, duration-agnostic lyrics-transcription module that maps grapheme and phoneme sequences to dense, frame-level aligned representations via unified multilingual tokenization and ConvNeXt-based \cite{woo2023convnext} refinement. To ensure precise temporal alignment across modalities, the model incorporates phase-aligned anisotropic positional infusion, selectively applying Rotary Positional Embeddings (RoPE) to temporally structured inputs such as video, audio, lyrics and transcription. Together, these components enable AudioGen-Omni to produce temporally synchronized, semantically coherent audio outputs with strong cross-modal integration and generalization capabilities. The primary contributions of this work are summarized as follows:
\begin{itemize}
\item To the best of our knowledge, AudioGen-Omni is the first unified framework capable of generating diverse audio types—including general audio, speech, and song—under flexible multimodal conditions, enabling precise audio-visual alignment.
\item A lightweight module maps raw grapheme or phoneme sequences to dense, frame-aligned representations without requiring phoneme duration supervision. It supports multilingual input with unified VoiceBPE tokenization and ConvNeXt-based refinement.
\item To enable cross-modal temporal resonance, phase-aligned anisotropic positional infusion selectively embeds rotational positional priors into temporally structured modalities—visual, audio, and aligned text like lyrics and transcription—reinforcing fine-grained synchrony across representations.
\end{itemize}
\section{Related Work}
\label{related_work}

\textbf{Video-to-Audio Synthesis.} Video-to-audio (V2A) \cite{luo2023diff, liu2023audioldm, luo2023diff, wang2024frieren} generation aims to synthesize meaningful audio signals that correspond to the visual content in a video. This technology finds broad applications, including enriching silent videos and enhancing multimedia content creation. A core challenge in V2A lies in the fact that visual data does not inherently contain audio information; instead, it provides indirect cues such as object movements, interactions, and environmental context. Successfully translating these visual signals into realistic and contextually appropriate audio requires sophisticated understanding and modeling of cross-modal relationships.

Recent advances \cite{lipman2022flow, rai2025egosonics} have employed deep learning techniques to better align visual and auditory modalities, pushing forward the quality of video-driven audio synthesis. Nonetheless, most existing solutions are tailored to specific audio categories—like environmental sounds, music, or speech—resulting in limited adaptability to diverse and complex audiovisual scenarios. Consequently, developing a unified, robust framework capable of flexibly generating various types of audio from visual inputs while maintaining semantic relevance and temporal coherence remains a significant research challenge.

\noindent\textbf{Video-to-Speech Synthesis.} Video-to-speech (V2S) \cite{zhang2025deepaudio, kim2025faces} synthesis represents a particularly intricate subset of V2A generation, as it involves producing intelligible speech synchronized with the speaker’s lip movements and contextual visual cues. While text-to-speech (TTS) systems have made remarkable progress in generating natural and expressive speech via neural vocoders and transformer-based architectures \cite{wang2023neural, du2024cosyvoice, chen2024f5, anastassiou2024seed}, V2S synthesis must infer speech content purely from visual input, without relying on text transcription. Although recent methods have improved lip-synchronized speech generation, they typically operate under controlled conditions and face difficulties adapting to the variability of real-world settings. Meanwhile, general V2A models capable of generating diverse audio types from video have demonstrated promising results but have not yet been effectively integrated into V2S frameworks. To advance V2S, there is potential in leveraging pretrained V2A models to go beyond lip-reading, incorporating richer visual information—such as facial expressions, gestures, and scene context—to generate more coherent, expressive, and contextually appropriate speech.

\noindent\textbf{Song Generation.} Early works. Jukebox \cite{dhariwal2020jukebox} pioneered full-song synthesis by cascading multi-scale VQ-VAEs and transformers, yet its control is limited to genre/artist tags and inference is slow.
Multi-stage pipelines. Melodist \cite{hong2003effect} and MelodyLM \cite{li2023ntire} decompose the task into text-to-MIDI, singing-voice synthesis, and vocal-to-accompaniment alignment. While they improve vocal quality, the multi-stage design complicates training/inference and their datasets are restricted to Mandarin pop.
Joint modeling. SongCreator \cite{lei2020nonpharmaceutical} employs a dual-sequence LM to jointly generate vocals and accompaniment, but lacks textual control and yields muffled vocals. Freestyle \cite{ning2006effect} focuses on rap generation given lyrics and beats, sacrificing melodic diversity. Yue \cite{yuan2017forecasting} scales data and parameters with a two-stage, track-decoupled LM, achieving strong results at high cost.
Commercial systems. Suno, Udio, and SeedMusic \cite{bai2024beyond} deliver high-fidelity song, yet remain closed-source and provide limited controllability.

\begin{figure*}[t]
\centering
\includegraphics[width=1\textwidth]{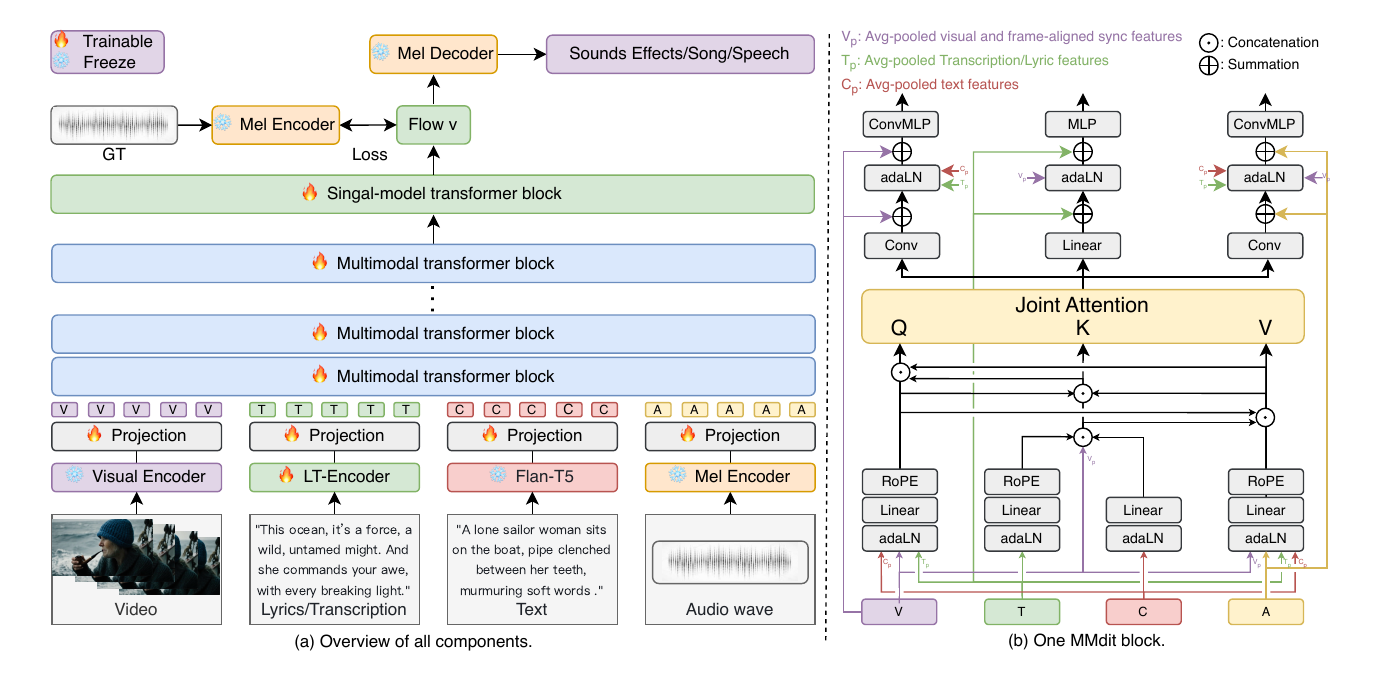}
\caption{Overview of the AudioGen-Omni  flow-prediction network. Video conditions, text conditions, lyric/transcript conditions and audio latents jointly interact in the multimodal transformer network.}
\label{fig:2}
\end{figure*}

\section{Method}

To generate high-quality audio, speech, music, or song from optional video and/or textual inputs within an end-to-end framework, we propose a multimodal architecture termed AudioGen-Omni. The primary objective of this approach is to effectively model the interactions among video, diverse audio types, and text modalities. To achieve this, we adopt the MM-DiT block design from SD3 \cite{esser2024scaling, cheng2025mmaudio} and integrate a series of audio-specific unimodal blocks inspired by FLUX \cite{labs2025flux}. This multimodal architecture enables adaptive attention to varying input modalities, thereby facilitating joint training on audio-visual and audio-text datasets.

\subsection{Automated Data Preprocessing Pipeline}

The effectiveness of AudioGen-Omni relies on a large-scale, diverse multimodal dataset encompassing text-to-audio/song/speech, video-to-audio/speech/song, and combined text-and-video-to-audio/speech/song pairs. This comprehensive dataset offers rich and flexible conditioning signals for model training.

\noindent\textbf{Descriptive Captions.} Utilizing Qwen-omni \cite{xu2025qwen2}, we automatically generate detailed textual descriptions that capture not only the acoustic content but also the prevailing mood and emotional dynamics of each audio sample.

\noindent\textbf{Speech Transcription.} Spoken segments are precisely transcribed using Whisper \cite{radford2023robust}, ensuring accurate phonetic and semantic representations across multiple languages and acoustic environments.

\noindent\textbf{Lyrics.} For musical content, lyrics are extracted and transcribed via FunASR \cite{gao2023funasr}, a robust Chinese-centric ASR toolkit, providing precise frame-level timing and punctuation to facilitate subsequent alignment and generation processes.

\subsection{Conditioning Encoders}

\noindent\textbf{Lyrics-Transcription Module.} 
Unlike prior non-autoregressive TTS systems that depend on pre-estimated phoneme durations, we propose a lightweight, duration-free lyrics-transcription module inspired by F5-TTS \cite{chen2024f5} and Ace-step \cite{gong2025ace}. This module directly maps raw grapheme or phoneme sequences into dense, frame-aligned representations. Non-Roman scripts are first converted to phonemes, followed by unified multilingual VoiceBPE tokenization. Learnable 768-dimensional embeddings are padded to the frame budget and masked at padding positions, enhanced with sinusoidal absolute positional encodings up to 4,000 positions, and refined through ConvNeXt-V2 blocks that respect the padding mask.

\noindent\textbf{Text Encoder.} 
We employ T5-Base \cite{raffel2020exploring}, pretrained on the Colossal Clean Crawled Corpus (C4), as the textual feature extractor. By unifying prompts, descriptions, and queries under a text-to-text framework, T5 produces robust 768-dimensional latent embeddings that serve as semantic anchors for downstream multimodal alignment and generation. Its strong generalization capacity reduces the need for task-specific tuning.

\noindent\textbf{Vision Encoder.} 
Visual features are extracted using ViT-bigG-14-QuickGELU from MetaCLIP \cite{ma2024mode}, pretrained on large-scale image-text datasets to yield domain-robust, fine-grained embeddings aligned with textual representations. To ensure temporal coherence, we integrate Synchformer \cite{iashin2024synchformer}, a Transformer-based audio-visual synchronization model that leverages sparse cues such as lip movements and phoneme timing, enabling precise alignment without dense supervision for applications including video generation, dubbing, and speech-driven animation.

\noindent\textbf{Audio Encoder.} 
Our audio encoder is based on the latent codec architecture from Kling-Foley \cite{wang2025kling}, an enhanced variant of the VQ-CTAP framework \cite{qiang2025vq} with improved reconstruction fidelity. The codec employs a Mel-spectrogram-based variational autoencoder (Mel-VAE), comprising an encoder, decoder, and discriminator. Input waveforms sampled at 44.1 kHz are encoded into latent embeddings at 43 Hz, achieving a temporal downsampling factor of 1024. By modeling a continuous latent distribution, this VAE attains higher representation capacity and reconstruction quality compared to discrete encoders, while maintaining compression efficiency.

\subsection{Input Strategies and Robustness}

To improve model robustness and adaptability to diverse input conditions, we adopt the following strategies:

\noindent\textbf{Multimodal Alignment.} 
By unfreezing all modalities and masking absent inputs, the model avoids the semantic lock-in inherent in text-frozen paradigms, enabling descriptive captions, transcription, lyrics, and video to jointly form a unified latent space. Shared projection layers and joint attention mechanisms facilitate unrestricted gradient flow, allowing low-resource modalities to leverage semantic information from richer modalities. This results in a modality-agnostic latent representation, permitting arbitrary subsets of conditioning inputs during inference without retraining. Furthermore, 24 FPS visual features ensure frame-level audio-visual synchronization without requiring computationally intensive test-time alignment.

\noindent\textbf{Variable-Length Training.} 
To support variable-length audio-visual generation with fine-grained temporal control, the original clip’s start time and duration are discretized into learnable per-second embeddings. These temporal embeddings are concatenated with global textual and visual features, fused with the diffusion timestep embedding via a shallow MLP, and incorporated into each transformer layer through adaptive layer normalization (AdaLN) \cite{perez2018film}, providing timing-aware global conditioning. During training, a length-based mask excludes padded frames from loss calculation, ensuring accurate gradient updates.

\subsection{Model Architecture}

\noindent\textbf{Joint Attention.} 
Drawing inspiration from Flux \cite{labs2025flux} and SD3 \cite{esser2024scaling}, we implement a joint attention mechanism to facilitate cross-modal information exchange. Specifically, query, key, and value representations from text, audio, and visual modalities are concatenated and processed via scaled dot-product attention \cite{shen2021efficient} over the combined sequence. This unified attention enables integrated cross-modal reasoning within a single operation. The output is subsequently partitioned according to the original modality structure, preserving modality-specific characteristics while enriching each with contextual information from other modalities.

\noindent\textbf{Phase-Aligned Anisotropic Positional Infusion (PAAPI).} 
Accurate temporal alignment across modalities is critical for coherent audiovisual synthesis. To address this, we propose Phase-Aligned Anisotropic Positional Infusion, a positional embedding strategy that selectively applies rotational positional encodings to temporally structured inputs—namely visual, audio, and temporally aligned textual streams such as lyrics and transcription—while maintaining isotropic embeddings in atemporal modalities. This anisotropic infusion enhances fine-grained temporal coherence by aligning phase-consistent positional information within the joint attention framework, as depicted in Figure \ref{fig:2}.

\begin{table*}[htbp]
\centering
\caption{Evaluation of audio generation methods on the VGGSound test set.}
\label{tab:1}
\resizebox{\textwidth}{!}{
\begin{tabular}{lccccccccc}
\toprule
\multirow{2}{*}{Method} & \multirow{2}{*}{Params} & \multicolumn{3}{c}{Distribution matching} & Audio quality & \multicolumn{2}{c}{Semantic align} & Temporal align & \\
\cmidrule(r){3-5} \cmidrule(r){6-6} \cmidrule(r){7-8} \cmidrule(l){9-9}
 &  & FD\textsubscript{PaSST}$\downarrow$ & FD\textsubscript{PANNs}$\downarrow$ & KL\textsubscript{PaSST}$\downarrow$ & IS$\uparrow$ & IB-score$\uparrow$ &  & DeSync$\downarrow$ & Time (s)$\downarrow$ \\
\midrule
ReWaS~\cite{jeong2025read} & 619M & 141.38 & 17.54 & 2.82 & 8.51 & 14.82 & & 1.062 & 15.97 \\
Seeing\&Hearing~\cite{xing2024seeing} & 415M & 219.01 & 24.58 & 2.30 & 8.58 & 33.99 & & 1.204 & 14.55 \\
V-AURA~\cite{viertola2025temporally} & 695M & 218.50 & 14.80 & 2.07 & 10.08 & 27.64 & & 0.654 & 16.55 \\
VATT~\cite{liu2024tell} & -- & 131.88 & 10.63 & 1.41 & 11.90 & 25.00 & & 1.195 & -- \\
Frieren~\cite{wang2024frieren} & 159M & 106.10 & 11.45 & 2.86 & 12.25 & 22.78 & & 0.851 & -- \\
FoleyCrafter~\cite{zhang2024foleycrafter} & 1.22B & 140.09 & 16.24 & 2.23 & 15.68 & 25.68 & & 1.225 & 1.67 \\
V2A-Mapper~\cite{wang2024v2a} & 229M & 84.57 & 8.40 & 2.56 & 12.47 & 22.58 & & 1.225 & -- \\
MMAudio-L-44.1kHz~\cite{cheng2025mmaudio} & 1.03B & \textbf{60.60} & \textbf{4.72}& \textbf{1.40} & \underline{17.40} & \textbf{33.22} & & \textbf{0.442} & 1.96 \\
Ours & 1.55B & \textbf{58.766} & \textbf{6.292} & \underline{1.556} & \textbf{21.521} & \underline{29.261} & & \underline{0.450} & 1.91 \\
\bottomrule
\end{tabular}
}
\end{table*}

\subsubsection{Global Conditioning} 
We construct a global conditioning vector shared across all Transformer layers by aggregating Fourier-encoded diffusion timesteps \cite{vaswani2017attention}, audio duration embeddings, and average-pooled visual and textual features. In contrast, the Lyric/Transcript representations provide localized temporal detail and are concatenated with Flan-T5 embeddings along the temporal dimension as part of the attention key. Following MMAudio, we note that although cross-modal attention facilitates interaction between visual and audio streams, the inherent soft aggregation may compromise alignment precision. To improve synchronization, we incorporate high-frame-rate (24 FPS) visual features extracted by the Synchformer encoder \cite{iashin2024synchformer}, which correlate strongly with audio events. These features are upsampled and integrated into the global conditioning vector to produce a frame-aligned conditioning signal. Both global and aligned features modulate the model through scale and bias parameters within adaptive layer normalization (AdaLN) layers.

\subsubsection{Conditional Flow Matching} 
During training, we employ conditional flow matching \cite{lipman2022flow, tong2023improving}. Given a condition \(C\) (e.g., text or video embedding), a noise vector \(x_0\) is sampled from a standard normal distribution. The model learns a velocity field \(v_\theta(t, C, x)\), and the training objective minimizes the discrepancy between the predicted velocity and the true flow velocity along the linear interpolation path, formalized as:

\begin{equation}
\label{e:cfm}
    \mathcal{L}_{\text{CFM}} = \mathbb{E}_{t, x_0, x_1, C} \left\| v_\theta(t, C, x_t) - u(x_t \mid x_0, x_1) \right\|^2,
\end{equation}

where \(x_t = (1 - t) x_0 + t x_1\), and \(u(x_t \mid x_0, x_1) = x_1 - x_0\). Here, \(t \in [0, 1]\) is the integration time, \(C\) is the condition (e.g., video and/or text), and \(x_t\) is a linearly interpolated point between noise and data. At inference time, we set \(t = 0.05\) and use Euler integration to map noise \(x_0\) to the final audio latent code.

\section{Experiments}
\noindent\textbf{Training Details.} We trained a model capable of generating 10-second audio, speech, or song outputs conditioned on multi-modal inputs.
The model has a total of 1.5 billion parameters, and the DiT model consists of 24 layers.
The training process uses the InverseLR optimizer with a base learning rate of 1e-5 and a weight decay of 0.001, along with a learning rate scheduler that incorporates exponential warm-up and decay phases. To improve inference stability, we maintain an exponential moving average of the model weights.
Training is conducted on eight clusters of NVIDIA H800 GPUs, each with 80GB of memory, requiring approximately 3000 GPU hours in total. The batch size is set to 128.
During inference, we perform 25 sampling steps using classifier-free guidance with a guidance scale of 4.5.

\noindent\textbf{Datasets.} We use VGGSound \cite{chen2020vggsound}, Pandas70M (approximately 4100 hours) \cite{chen2024panda}, and InterVid \cite{wang2023internvid} (approximately 1900 hours) as audio-text-visual datasets for training.
For audio-text training, we use AudioCaps \cite{kim2019audiocaps} (approximately 128 hours, manually captioned), Clotho \cite{drossos2020clotho} (approximately 31 hours, manually captioned), LibriTTS \cite{zen2019libritts} (approximately 585 hours), LJ Speech \cite{ren2019fastspeech} (approximately 24 hours), and WavCaps \cite{mei2024wavcaps} (approximately 7,600 hours, automatically captioned from metadata).
The song-lyrics training dataset is collected from online sources, totaling approximately 1,000 hours.

\subsection{Metrics}
We evaluate audio generation using four criteria: distribution similarity, audio fidelity, semantic coherence, and temporal alignment, as shown in Table \ref{tab:1}. For speech-specific assessment, we employ UTMOS~\cite{saeki2022utmos}, DNSMOS~\cite{reddy2021dnsmos}, and Word Error Rate (WER) to measure intelligibility, with results summarized in Table \ref{tab:2}. We also compute the Speaker Embedding Cosine Similarity (SECS) between synthesized and target speech on the LRS3 test set to evaluate speaker consistency, as reported in Table \ref{tab:3}.

\begin{table*}[htbp]
\centering
\caption{Evaluation of speech generation methods on both LRS3 and LRS2 test datasets.}
\label{tab:2}
\resizebox{\textwidth}{!}{
\begin{tabular}{llcccccccc}
\toprule
\multirow{2}{*}{Method} & \multirow{2}{*}{Steps} & \multicolumn{4}{c}{\textbf{LRS3-TED}} & \multicolumn{4}{c}{\textbf{LRS2-BBC}} \\
\cmidrule(r){3-6} \cmidrule(l){7-10}
& & UTMOS$\uparrow$ & DNSMOS$\uparrow$ & RMSE$_{f0}$$\downarrow$ & WER$\downarrow$ & UTMOS$\uparrow$ & DNSMOS$\uparrow$ & RMSE$_{f0}$$\downarrow$ & WER$\downarrow$ \\
\midrule
Ground Truth & -- & 3.545 & 2.582 & -- & 2.29 & 3.013 & 2.256 & -- & 8.93 \\
\midrule
\multicolumn{10}{l}{\textit{Audio-driven speaker embedding}} \\
SVTS~\cite{mira2022svts} & -- & 1.283 & 1.860 & 56.929 & 84.98 & 1.387 & 1.434 & 53.475 & 83.38 \\
Intelligible~\cite{choi2023intelligible} & -- & 2.702 & 2.395 & 39.377 & \underline{29.60} & 2.331 & 2.000 & \underline{41.233} & 39.53 \\
\midrule
\multicolumn{10}{l}{\textit{Video-driven speaker embedding}} \\
LTBS~\cite{kim2024let} & -- & 2.417 & 2.361 & 40.006 & 84.08 & 2.288 & 2.174 & 43.653 & 94.25 \\
DiffV2S~\cite{choi2023diffv2s} & 1000 & 3.058 & 2.558 & 40.893 & 41.07 & 2.945 & 2.363 & 44.414 & 54.86 \\
Faces2Voices \cite{kim2025faces} & 1000 & \textbf{3.993} & \underline{2.759} & \underline{38.928} & \underline{30.37} & \textbf{3.881} & \underline{2.552} & \underline{43.702} & \underline{39.05} \\
Ours & 25 & \underline{3.982} & \textbf{3.782} & \textbf{37.525} & \textbf{17.56} & \underline{}{3.842} & \textbf{3.767} & \textbf{42.902} & \textbf{17.75} \\
\bottomrule
\end{tabular}
}
\end{table*}

\subsection{Main Results}
\subsubsection{Audio Generation}

\textbf{Distribution Similarity.} To assess how closely the distribution of generated audio matches that of real audio, we compute the \textit{Fréchet Distance (FD)} and the \textit{Kullback--Leibler (KL) divergence} using features extracted from multiple pretrained models. For FD, we adopt three embedding models: PaSST~\cite{koutini2021efficient} (FD\textsubscript{PaSST}) and PANNs~\cite{kong2020panns} (FD\textsubscript{PANNs}). Note that PaSST operates at 32~kHz, while PANNs works at 16~kHz. Additionally, PaSST and PANNs generate global representations.

For KL divergence, we follow the implementation of Liu et al.~\cite{liu2023audioldm} as classifiers to compute the class distribution differences between generated and real samples.

\noindent\textbf{Audio Fidelity.} We evaluate perceptual quality without requiring ground-truth audio by using the \textit{Inception Score (IS)}~\cite{girdhar2023imagebind}. Following Wang et al.~\cite{viertola2025temporally}, we use PANNs as the classifier to calculate the IS.

\noindent\textbf{Semantic Coherence.} To measure how well the generated audio semantically aligns with the input video, we use ImageBind~\cite{girdhar2023imagebind} to extract visual and audio embeddings. We then compute the average cosine similarity between the modalities as our \textit{IB-score}, following Viertola et al.~\cite{viertola2025temporally}.

\begin{figure}[t]
\centering
\includegraphics[width=1.1\columnwidth]{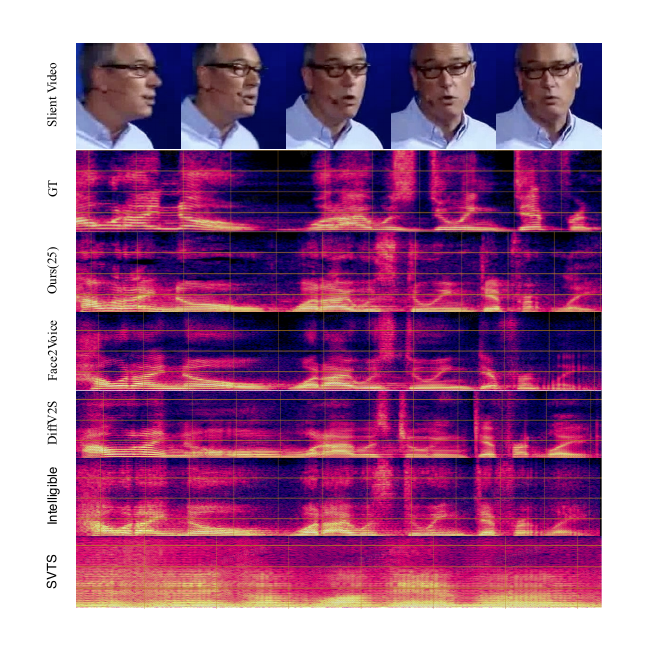} 
\caption{Mel-spectrogram visualization compared with Ground Truth (GT) speech demonstrates that the proposed method successfully captures both precise and expressive variations in fundamental frequency, along with facial expressions that are closely synchronized over time.}
\label{fig:3}
\end{figure}

\noindent\textbf{Temporal Alignment. } To evaluate audio-visual synchronization, we adopt the \textit{DeSync} score predicted by Synchformer~\cite{iashin2024synchformer}, which estimates the temporal misalignment (in seconds) between audio and video. Unlike Viertola et al.~\cite{viertola2025temporally}, who evaluate with 2.56-second clips (shorter than Synchformer's 4.8-second context window), we use 8-second clips. We extract two crops (first 4.8s and last 4.8s) and average the DeSync values to obtain a more robust synchronization estimate.

\begin{table}[t]
\centering
\caption{SECS evaluation results on LRS3 test set.}
\begin{tabular}{lcccc}
\toprule
Method & LTBS & DiffV2S & Faces2Voices (1000) & Ours (25) \\
\midrule
GE2E$\uparrow$   & 0.609 & 0.621 & \underline{0.650} & \textbf{0.691} \\
VoxSim$\uparrow$ & 0.399 & 0.433 & \underline{0.494} & \textbf{0.527}\\
\bottomrule
\end{tabular}
\label{tab:3}
\end{table}

\subsubsection{Speech Generation}
\noindent\textbf{Speech Objective Evaluation.} We evaluate the quality of the generated speech using two widely adopted perceptual audio quality assessment models: UTMOS  and DNSMOS. Additionally, we compute the root mean square error of F0 (RMSEf0) to measure pitch accuracy, and the Word Error Rate (WER) to assess speech intelligibility. WER is calculated by transcribing the generated speech using the Whisper 3.0 and comparing it to the ground-truth transcription.

Our model outperforms existing VTS systems on both the LRS3 and LRS2 datasets, demonstrating its effectiveness in reducing the modality gap between video and speech. Notably, our method even surpasses ground-truth audio in terms of UTMOS and DNSMOS scores, which can be attributed to the generation of clean speech without background noise, in contrast to real-world recordings.

\noindent\textbf{Analysis on Speaker Similarity.} We further evaluate whether the video-driven speaker embeddings can effectively capture speaker identity. To this end, we compute the Speaker Embedding Cosine Similarity (SECS) between the synthesized and target speech on the LRS3 test set. Speaker embeddings are extracted using two different models: GE2E \cite{wan2018generalized}, a standard speaker verification model, and VoxSim \cite{ahn2024voxsim}, which is specifically designed to measure perceptual voice similarity.

As shown in Table \ref{tab:3}, our method achieves the highest SECS scores across both embedding models, demonstrating that video-driven embeddings in our approach more accurately preserve speaker characteristics compared to existing methods.

\noindent \textbf{Mel-spectrogram Visualization.} For a more intuitive comparison with baseline methods, we visualize the generated speech using mel-spectrograms alongside the ground-truth audio. As shown in Figure \ref{fig:3}, the mel-spectrogram produced by our model closely matches the ground-truth counterpart, accurately capturing fine acoustic details and harmonic structures.

Moreover, our approach effectively enhances prosody by leveraging visual features, as evidenced by dynamic variations in the fundamental frequency (F0) that correspond with abrupt facial expression changes.

\section{Conclusion}
We propose AudioGen-Omni, a unified multimodal diffusion transformer that generates high-fidelity audio, speech, and song synchronized with input video. Leveraging large-scale video-text-audio training data, it employs a unified lyrics-transcription encoder and a novel joint attention with phase-aligned positional infusion to ensure precise cross-modal alignment. By unfreezing all modalities and masking missing inputs, AudioGen-Omni overcomes limitations of text-frozen models, enabling flexible conditioning and strong generalization. It achieves state-of-the-art results on multiple audio generation tasks with efficient inference, laying the groundwork for future extensions including video generation.
\bibliography{iclr2025_conference}
\bibliographystyle{iclr2025_conference}


\end{document}

%% file: math_commands.tex
\usepackage{amsmath,amsfonts,bm}

\def\eqref#1{equation~\ref{#1}}

\def\1{\bm{1}}

\DeclareMathAlphabet{\mathsfit}{\encodingdefault}{\sfdefault}{m}{sl}
\SetMathAlphabet{\mathsfit}{bold}{\encodingdefault}{\sfdefault}{bx}{n}

